\documentclass[fleqn,usenatbib]{mnras}
\usepackage[T1]{fontenc}
\usepackage{ae,aecompl}
\usepackage[dvipdfmx]{graphicx}	
\usepackage{multirow}
\usepackage{amsmath}
\usepackage{amssymb}
\usepackage{color,ulem}

\def\Msun{M_{\odot}}
\def\num{\nu_{\rm m}}
\def\nua{\nu_{\rm a}}
\def\nuc{\nu_{\rm c}}
\def\epe{\varepsilon_{\rm e}}
\def\epb{\varepsilon_{\rm B}}

\newcommand{\be}{\begin{equation}}
\newcommand{\ee}{\end{equation}}

\title[AT 2022cmc]{Synchrotron Afterglow Model for AT 2022cmc: Jetted Tidal Disruption Event or Engine-Powered Supernova?}

\author[Matsumoto \& Metzger]{Tatsuya Matsumoto$^{1}$\thanks{E-mail: tm3238@columbia.edu} and Brian D. Metzger$^{1,2}$\\
$^{1}$Department of Physics and Columbia Astrophysics Laboratory, Columbia University, Pupin Hall, New York, NY 10027, USA\\
$^{2}$Center for Computational Astrophysics, Flatiron Institute, 162 5th Ave, New York, NY 10010, USA\\
}

\pubyear{2023}

\begin{document}
\label{firstpage}
\pagerange{\pageref{firstpage}--\pageref{lastpage}}
\maketitle

\begin{abstract}
AT 2022cmc is a luminous optical transient ($\nu L_{\nu} \gtrsim 10^{45}$ erg s$^{-1}$) accompanied by decaying non-thermal X-rays (peak duration $t_{\rm X} \lesssim$ days and isotropic energy $E_{\rm X,iso} \gtrsim 10^{53}$ erg) and a long-lived radio/mm synchrotron afterglow, which has been interpreted as a jetted tidal disruption event (TDE). Both an equipartition analysis and a detailed afterglow model reveals the radio/mm emitting plasma to be expanding mildly relativistically (Lorentz factor $\Gamma \gtrsim\,few$) with an opening angle $\theta_{\rm j}\simeq0.1$ and roughly fixed energy $E_{\rm j,iso} \gtrsim few \times 10^{53}$ erg into an external medium of density profile $n \propto R^{-k}$ with $k \simeq 1.5-2$, broadly similar to that of the first jetted TDE candidate Swift J1644+57 and consistent with Bondi accretion at a rate $\sim 10^{-3}\dot{M}_{\rm Edd}$ onto a $10^{6}M_{\odot}$ black hole before the outburst. The rapidly decaying optical emission over the first days is consistent with fast-cooling synchrotron radiation from the same forward shock as the radio/mm emission, while the bluer slowly decaying phase to follow likely represents a separate thermal emission component.  Emission from the reverse shock may have peaked during the first days, but whose non-detection in the optical band places an upper bound $\Gamma_{\rm j} \lesssim 100$ on the Lorentz factor of the unshocked jet.  Although a TDE origin for AT 2022cmc is indeed supported by some observations, the vast difference between the short-lived jet activity phase $t_{\rm X} \lesssim$ days relative to the months-long thermal optical emission, also challenges this scenario. A stellar core-collapse event giving birth to a magnetar or black hole engine of peak duration $\sim 1$ day offers an alternative model also consistent with the circumburst environment, if interpreted as a massive-star wind.
\end{abstract}


\section{Introduction}
\label{sec:introduction}

AT 2022cmc was discovered as a luminous, rapidly evolving transient by the Zwicky Transient Facility \citep{Andreoni+2022,Pasham+2022}.  A potential host galaxy was identified at redshift $z \simeq 1.19$ (luminosity distance $d_{\rm L}=8.44\,\rm Gpc$), which we adopt hereafter.  The optical light curve of AT 2022cmc rose to a peak luminosity $\nu L_{\nu} \sim 10^{46}$ erg s$^{-1}$ within a day after its discovery, before decaying by several magnitudes over the course of a week and settling into a slowly-declining ``plateau'' of luminosity $\sim10^{45}\,\rm erg\,s^{-1}$ lasting at least a couple of months (hereafter, all times are as measured in the observer frame, unless stated otherwise).  While the early rapidly evolving emission was red in color, the later plateau instead exhibited a featureless blue continuum spectrum with a blackbody temperature of $\simeq(2-4)\times10^{4}\,\rm K$ and low levels of optical polarization (consistent with zero; \citealt{Cikota+2023}).  Multi-wavelength followup starting $\simeq5\,\rm days$ after the discovery resulted in the detection of time-variable non-thermal X-ray emission, whose luminosity was rapidly decaying $L_{\rm X,iso}\simeq3\times10^{47}{\,\rm erg\,s^{-1}\,}(t/5\,\rm days)^{-2}$, corresponding to a total radiated energy $E_{\rm X,iso} \gtrsim 10^{53}$ erg.  Observations at radio/mm wavelengths revealed a highly luminous time-evolving source with a synchrotron self-absorbed (SSA) spectrum. 

\cite{Andreoni+2022,Pasham+2022} interpreted AT 2022cmc as a tidal disruption event (TDE, \citealt{Hills1975,Rees1988}) of a star by a supermassive black hole (SMBH) accompanied by a relativistic jet (e.g., \citealt{Giannios&Metzger2011}; see \citealt{DeColle&Lu2020} for a review of jetted TDEs), in part due to the similarity of its non-thermal X-ray and radio emission to previous jetted TDE candidates, particularly the well-studied Swift J1644+57 \citep{Bloom+2011,Burrows+2011,Levan+2011,Zauderer+2011,Wiersema+2012}.  A TDE association for AT 2022cmc is also supported by the similarity of its blue optical plateau phase with the thermal emission of other better-confirmed TDEs \citep{vanVelzen+2019,hammerstein+2023}, including the second jetted TDE candidate Swift J2058+05 \citep{Cenko+2012,Pasham+2015,Wiersema+2020}. However, the lack of a confirmed location of AT 2022cmc in the nucleus of the host galaxy (due to the transient outshining the host light), as well as the short peak timescale of the X-ray emission ($\lesssim$ days)\footnote{It would appear unlikely that the jet was launched well before the epoch of the optical discovery. Such a temporal offset would flatten the X-ray light curve up to roughly the same duration as the offset \citep{Tchekhovskoy+2014}, contrary to the observed power-law decay in AT 2022cmc if $t=0$ is taken at optical discovery.} compared to other jetted TDE candidates or to the predicted fallback timescale of the disrupted stellar material ($\simeq\,\rm month$ for typical SMBH masses $\gtrsim 10^{6}M_{\odot}$), leaves open the possibility for other explanations, particularly the birth of a central engine following the core-collapse of a star \citep{Quataert&Kasen2012,Nakauchi+2013,Metzger+2015d,Perna+18}.

The prototypical jetted TDE candidate Swift J1644+57 exhibited highly time-variable hard X-ray emission lasting roughly $t_{\rm X} \sim$ 10 days after the X-ray trigger, followed by a power-law decay $L_{\rm X} \propto t^{-5/3}$ similar to the expected late-time mass fallback rate from a TDE \citep{Bloom+2011,Burrows+2011,Levan+2011}, which abruptly terminated around $t \simeq 500$ days \citep{Zauderer+2013}, possibly as a result of a late-time state transition within the disk (e.g., \citealt{Tchekhovskoy+2014}); a remarkably similar X-ray drop was seen also in Swift J2058+05 \citep{Pasham+2015}.  The time-lag associated with reverberation mapping of the X-ray spectrum was found to support a SMBH mass of $\sim 10^{6}M_{\odot}$ \citep{Kara+2016,Lu+2017}.

As predicted by \citet{Giannios&Metzger2011} just months prior to Swift J1644+57, the onset of the X-ray emission was accompanied by luminous synchrotron radio emission \citep{Zauderer+2011,Berger+2012,Zauderer+2013,Eftekhari+2018,Cendes+2021}, created as the relativistic jet material collided with the sub-parsec scale circumnuclear medium surrounding the (nominally previously quiescent) SMBH.  Analysis of the radio emission showed the radiated plasma to be expanding relativistically, with a Lorentz factor $\Gamma\gtrsim few$ (e.g., \citealt{Zauderer+2011,Metzger+2012,Berger+2012,BarniolDuran&Piran2013}).  Although the long-lasting radio emission is well-modeled as originating from the forward shock (FS), a temporal break in the radio light curve around $\sim 2t_{\rm X}$ was interpreted as evidence for an early reverse shock (RS) phase \citep{Metzger+2012}.  The radio light curves also exhibited a re-brightening at $t \simeq 1$ month, implying either substantial late-time energy injection into the FS \citep{Berger+2012} and/or complex angular structure of the jet (e.g., a moderately relativistic ``sheath'' surrounding a high-$\Gamma$ jet core; \citealt{Tchekhovskoy+2014,Mimica+2015,Generozov+2017,Lu+2017}).

The mechanisms responsible for creating relativistic jets in TDEs remain unclear (e.g., \citealt{DeColle&Lu2020} for a review). The discovery of just three jetted TDEs over the past decade$-$Swift J1644+57, Swift J2058+05, and Swift J1112-82 \citep{Brown+2015}$-$implies a rate of on-axis jetted TDEs as small as $\sim0.01\,\rm Gpc^{-3}\,yr^{-1}$ \citep{Alexander+2020,DeColle&Lu2020}. Allowing for a beaming correction $\sim 100$ for the jetted X-rays, this implies that only a small fraction $\sim 1\%$ of TDEs create powerful jets; this low jetted TDE fraction is also supported by the lack of radio emission from off-axis jets in the majority of optical and (thermal) X-ray-selected TDEs (\citealt{Bower+2013,vanVelzen+2013,Generozov+2017,Matsumoto&Piran2021b}; however, see \citealt{Horesh+2021,Perlman+2022,Cendes+2022b,Sfaradi+2022,Matsumoto&Piran2022}).  The creation of prompt relativistic jets in TDEs via the Blandford-Znajek process \citep{Blandford&Znajek1977} may require special conditions, such as progenitor stars with atypically strong magnetic fields (e.g., \citealt{Giannios&Metzger2011,Bradnick+2017}) or a pre-existing active galactic nucleus (AGN) disk threaded by a sufficient magnetic flux (e.g., \citealt{Tchekhovskoy+2014,Kelley+2014}).

In this paper we model the radio and early-time optical emission from AT 2022cmc within the framework of an afterglow model \citep{Giannios&Metzger2011,Metzger+2012}, in order to constrain the properties of the relativistic jet from this event and the density profile of the surrounding gaseous medium.  In contrast to \cite{Pasham+2022}, we assume the early rapidly-declining X-rays arise from emission internal to the jet, as was the interpretation for Swift J1644+57 (e.g., \citealt{Lu&Kumar2016,Crumley+2016}) and is supported by the rapid X-ray variability time $\sim10^3\,\rm s$.  We likewise follow \citet{Andreoni+2022} in assuming the blue optical wavelength plateau phase to be thermal emission unrelated to the jet (e.g., originating from the photosphere of the TDE envelope/accretion disk, or the supernova (SN) ejecta in core collapse scenarios).

We begin in \S \ref{sec:equipartition} by applying basic equipartition arguments to constrain in relatively agnostic terms the properties of the radio emitting region, which we identify as gas behind the FS.  Then in \S \ref{sec:lightcurve} we present a detailed synchrotron afterglow light curve calculation, which can explain the early optical emission phase within the same FS model as the radio/mm emission (\S \ref{sec:preliminary}).  In \S \ref{sec:discussion} we discuss some implications of our findings, such as the prediction of an additional RS emission component.  We also address the physical origin of AT 2022cmc, comparing the TDE and core collapse scenarios, and their respective implications for the external medium probed by the afterglow.  We summarize our conclusions in \S \ref{sec:summary}.

\section{Equipartition Analysis}
\label{sec:equipartition}

\begin{table*}
\begin{center}
\caption{Results of the equipartition analysis (under the assumption of an on-axis viewed emitter) for two scenarios for the jet opening angle of $\theta_{\rm j}=0.1$ and $\theta_{\rm j}=1/\Gamma$. The quantities $\Gamma$, $R_{\rm eq}$, $E_{\rm eq}$, $E_{\rm tot}$, and $N_{\rm eq}$ are the (post-shock) bulk Lorentz factor, radius, energy, outflow's total energy, and number of emitting particles, respectively. Note $E_{\rm eq}$ here includes only that of the emitting electrons and the magnetic field, which are assumed to be comparable (i.e., $\epe\simeq\epb$), and $E_{\rm tot}$ includes the energy of hot proton as well taking into account for the deviation from the equipartiton with $\varepsilon_{\rm e}=0.1$ and $\varepsilon_{\rm B}=10^{-3}$ (see also the main text). The observation epochs for which rich data gives relatively secure $\nua$ and $\num$ are shown in boldface.}
\label{table:equipartition}
\begin{tabular}{cccc|ccccc|ccccc}
\hline
\multicolumn{4}{c|}{Observables}&\multicolumn{5}{c|}{Narrow jet ($\theta_{\rm j}=0.1$)}&\multicolumn{5}{c}{Wide jet ($\theta_{\rm j}=1/\Gamma$)}\\
\hline
$t$&$F_{\rm p}$&$\nu_{\rm a}$&$\nu_{\rm m}$&$\Gamma$&$R_{\rm eq}$&$E_{\rm eq}$&$E_{\rm tot}$&$N_{\rm eq}$&$\Gamma$&$R_{\rm eq}$&$E_{\rm eq}$&$E_{\rm tot}$&$N_{\rm eq}$\\
$[\rm day]$&[mJy]&[GHz]&[GHz]&&[cm]&[erg]&[erg]&&&[cm]&[erg]&[erg]\\
\hline
5.1&17.2&1.1e+02&1.2e+03&3.1&1.1e+17&1.0e+49&1.4e+50&8.3e+51&1.7&2.6e+16&2.0e+49&2.6e+50&3.0e+52\\
7.0&12.7&8.2e+01&7.1e+02&3.0&1.3e+17&1.1e+49&1.5e+50&9.6e+51&1.6&3.1e+16&2.1e+49&2.7e+50&3.6e+52\\
\bf 11.6&6.3&5.5e+01&3.6e+02&2.6&1.6e+17&9.5e+48&1.3e+50&1.1e+52&1.4&3.4e+16&2.0e+49&2.5e+50&4.1e+52\\
15.5&5.0&4.7e+01&2.2e+02&2.5&2.0e+17&9.8e+48&1.3e+50&1.3e+52&1.4&3.9e+16&2.1e+49&2.6e+50&5.0e+52\\
\bf 20.4&4.3&4.0e+01&1.5e+02&2.4&2.4e+17&1.1e+49&1.5e+50&1.6e+52&1.3&4.6e+16&2.3e+49&2.9e+50&6.1e+52\\
27.8&4.3&3.5e+01&1.2e+02&2.2&2.8e+17&1.4e+49&1.9e+50&2.1e+52&1.3&5.2e+16&3.0e+49&3.8e+50&8.3e+52\\
\bf 45.3&3.1&2.6e+01&1.2e+02&1.9&3.2e+17&1.8e+49&2.4e+50&2.8e+52&1.2&5.4e+16&3.7e+49&4.6e+50&1.0e+53\\
\hline
\end{tabular}
\end{center}
\end{table*}

\begin{figure}
\begin{center}
\includegraphics[width=85mm, angle=0]{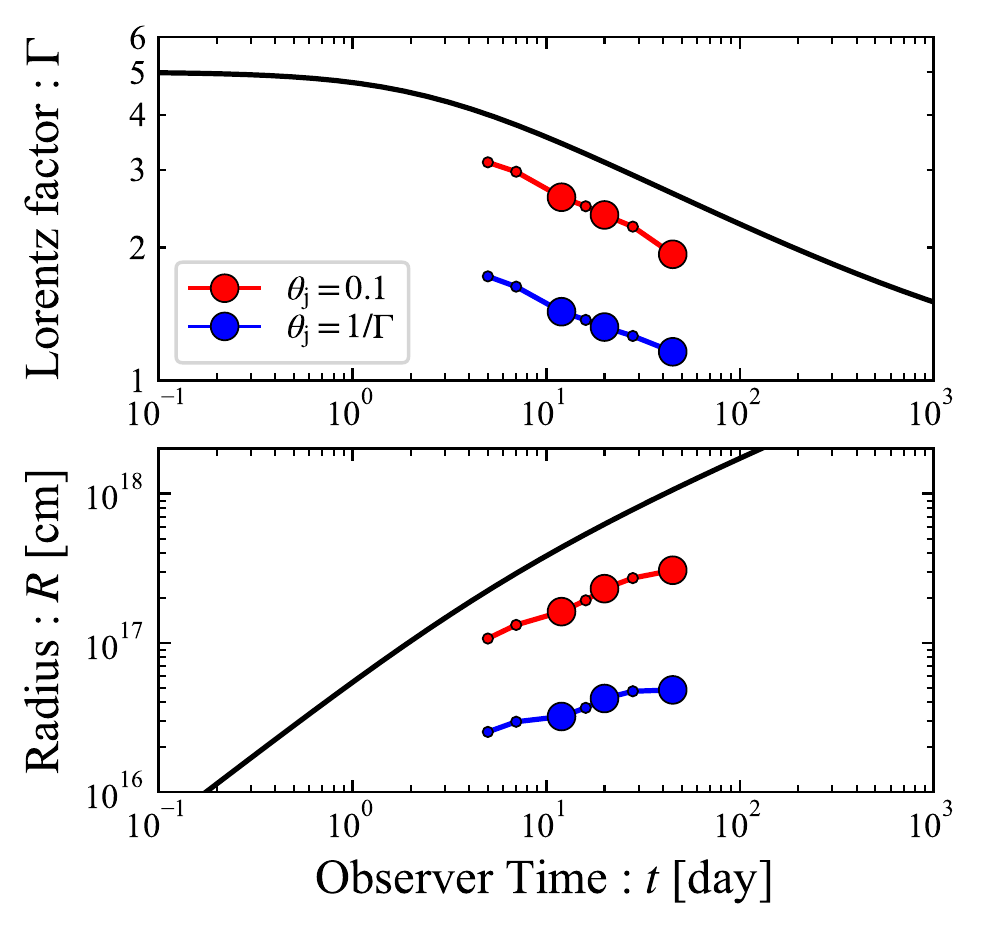}
\caption{The Lorentz factor (top) and radius (bottom) of the radio emitting region obtained by equipartition analysis. The red and blue points show the results for the models of $\theta_{\rm j}=0.1$ (narrow jet) and $\theta_{\rm j}=1/\Gamma$ (wide jet), respectively. Large dots represent the epochs for which the break frequencies are best determined due to a good spectral coverage. Black curves show the evolution of the quantities of our afterglow model (see \S \ref{sec:lc} for details).}
\label{fig:equipartition}
\end{center}
\end{figure}

The SSA radio spectrum of AT 2022cmc allows us to estimate the radius and Lorentz factor of the radio emitting region using the equipartition method \citep{Chevalier1998,BarniolDuran+2013}.
Using the radio spectra provided by \cite{Andreoni+2022} (their Extended Data Fig.~1), we carried out the equipartition analysis within the framework of \cite{Matsumoto&Piran2022} for an assumed on-axis jet orientation.  We adopted two scenarios for the geometry of the emitting region: one assumes a narrowly collimated jet with a constant half opening-angle $\theta_{\rm j}=0.1$, a choice motivated by observations of jets in AGN (e.g., \citealt{Pushkarev+2017}) and from modeling of Swift J1644+57 \citep{Metzger+2012}.  In this scenario, as the jet decelerates, we only receive emission from a region of fixed solid angle $\pi \theta_{\rm j}^{2}$ (we will see the size of the beaming cone $=1/\Gamma$ is larger than $\theta_{\rm j}=0.1$, where $\Gamma$ is the Lorentz factor of the emitting shocked gas).
The second model assumes a wider jet whose opening angle $\theta_{\rm j}=1/\Gamma$ grows as the jet decelerates.

\begin{figure}
\begin{center}
\includegraphics[width=85mm, angle=0]{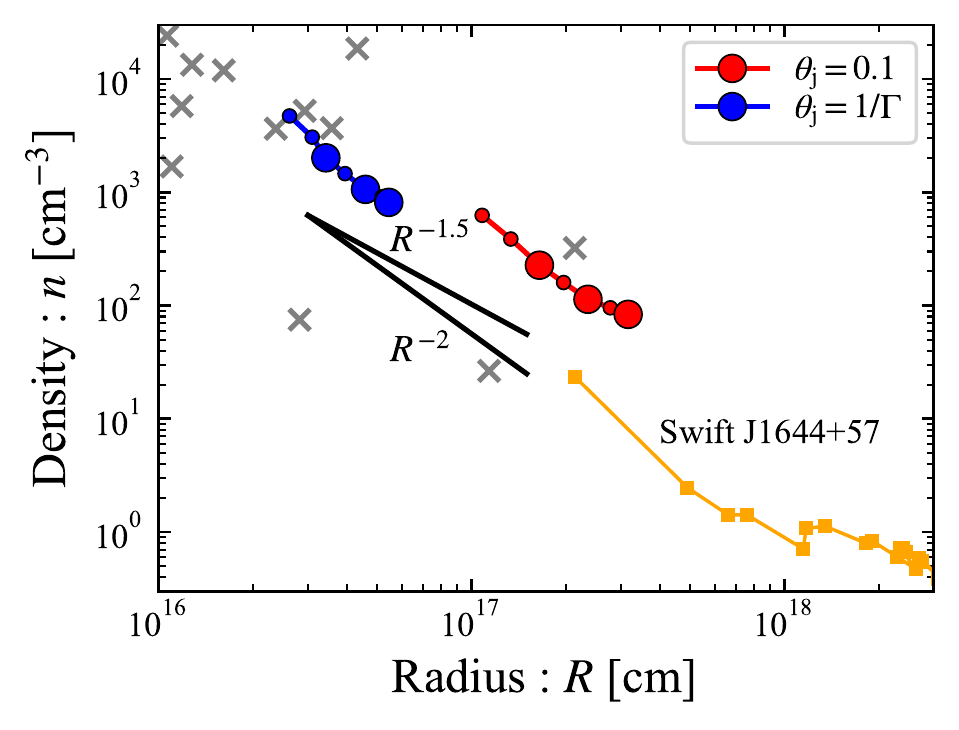}
\caption{Density profile of the external medium reconstructed by the equipartition method. The orange points represent the profile for Swift J1644+57 taken from \citealt{Eftekhari+2018}. For comparison, gray crosses show the inferred circumstellar-medium density of radio SNe (\citealt{Chevalier1998,Chevalier&Fransson2006}, as presented in \citealt{Bright+2022}).}
\label{fig:profile}
\end{center}
\end{figure}

Table~\ref{table:equipartition} shows the results of the equipartition analysis applied to each observational epoch for the two models. Since the spectral fitting by \cite{Andreoni+2022} finds that the SSA frequency is always smaller than the characteristic frequency, $\nua<\num$, the spectrum peaks at $\num$. For the earliest two epochs, the spectral peak $F_{\rm p}$ is not captured by the frequencies of the observations in \cite{Andreoni+2022}, so we obtain them by extrapolating their spectral fits assuming a sharp spectral break, which slightly overestimates the peak flux relative to what would be obtained assuming a smooth break. We also remark that the radio/mm data is limited in particular at mm wavelengths, which may allow different spectral fits.\footnote{The equipartition quantities weakly depend on the peak flux and frequency. For instance, the quantities in Figs.~\ref{fig:equipartition} and \ref{fig:profile} obey $\Gamma\propto F_{\rm p}^{1/5}\nu_{\rm p}^{-2/15}$, $R_{\rm eq}\propto F_{\rm p}^{2/5}\nu_{\rm p}^{-4/15}$, and $n\propto F_{\rm p}^{-3/5}\nu_{\rm p}^{11/15}$. Hence our result is also insensitive to the observables as long as the spectral fit of \cite{Andreoni+2022} is reasonably correct.} Nevertheless, our results provide a reasonable physical scenario. Fig.~\ref{fig:equipartition} shows the Lorentz factor and radius of the emitting region as a function of time.  For both models, the emitting region is found to be mildly relativistic with $\Gamma\simeq1.5-3$ and in a deceleration stage during the observations.  The Lorentz factor and radius of the narrow-jet ($\theta_{\rm j}=0.1$) model are larger than in the wide-jet model ($\theta_{\rm j}=1/\Gamma$) model because the smaller opening-angle requires a larger emitting region to reproduce the observed radio flux.

The equipartition energy $E_{\rm eq}$, which reflects only the energies (beaming corrected) of emitting electrons and magnetic field, remains almost constant in time $E_{\rm eq}\simeq10^{49}\,\rm erg$ for both scenarios. This implies that the emitting region does not undergo energy injection (or the injection is energetically negligible) at times $\gtrsim5\,\rm days$ covered by the radio data. This is despite the fact that the jet (as probed by the variable X-ray emission) does remain active for this period (though the total radiated X-ray energy is dominated by early times; see the next paragraph).  By making the ``equipartition'' assumption, the radius/energy are determined taking the energy of the emitting electrons and the magnetic field to be comparable; while deviations from equipartition do not modify the implied radius and the external density profile significantly, they do increase the required energy considerably \citep{BarniolDuran+2013}, a point to which we shall return in \S \ref{sec:preliminary}. The number of emitting particles $N_{\rm eq}$ increases weakly in both scenarios.

The temporal evolution of $E_{\rm eq}$ and $N_{\rm eq}$, as well as the delayed timing of the radio emission relative to the short-lived X-ray emission (taken as a proxy for the jet activity) implicates the radio emission as originating from the FS, i.e., the shocked external medium, as in Swift J1644+57 (e.g., \citealt{Metzger+2012,Berger+2012}).  In the case of an adiabatic FS, the energy of the shocked gas remains roughly conserved (absent ongoing energy injection) as the number of swept-up particles increases monotonically.  The absence of substantial energy injection is supported by the rapidly declining X-ray luminosity, $L_{\rm X}\propto t^{-2}$, consistent with most of the energy being injected into the blast wave $E \propto \int L_{\rm X}dt$ occurring prior to the start of the radio observations.  In contrast, if the emitting region were the RS passing back through the jet material, the energy of the emitting electrons should increase as more and more particles are swept up.  Contrary to \cite{Andreoni+2022}, we therefore disfavor the RS as the source of the observed radio/mm emission.

Fig.~\ref{fig:profile} depicts the density profile of the external medium ($n \equiv \rho/m_{\rm p}$, where $\rho$ is the mass density and $m_{\rm p}$ the proton mass) under the assumption of a FS origin for the radio emission. The profile for both scenarios is well-described by a power-law in radius $R$, 
\begin{align}
n(R)=n_{17}\left(\frac{R}{10^{17}\,\rm cm}\right)^{-k},
    \label{eq:profile}
\end{align}
where $n_{17}$ is the density at $R = 10^{17}$ cm.  The slope implied by the data $k\simeq1.5-2$ (black lines in Fig.~\ref{fig:profile}), is similar to that implied by modeling the early radio emission from Swift J1644+57 \citep{Metzger+2012,Berger+2012} and that from some optical TDEs \citep[e.g., AT 2019dsg,][]{Cendes+2021b,Matsumoto+2022}.  The density normalization is a factor of at least a few higher in AT 2022cmc than in Swift J1644+57 on a comparable radial scale $\sim 10^{17}$ cm.

\section{Light Curve Model}
\label{sec:lightcurve}

Building on the results of the equipartition analysis, we now construct a detailed synchrotron FS light curve model for AT 2022cmc.  We begin with some preliminary considerations, which motivate various aspects of the model, such as the total jet energy and the narrow vs. wide-jet scenario, and which argue in favor of a FS origin for the early optical emission phase.

\subsection{Preliminary Considerations}
\label{sec:preliminary}

The distinct shape and red colors of the early optical emission peaking at $t \simeq1\,\rm day$ \citep{Andreoni+2022}, suggest a separate origin from the bluer thermal plateau phase to follow.  Emission from both the FS and RS peaks when the RS crosses the unshocked-jet material  \citep{Sari&Piran1999b,Kobayashi2000,Chevalier&Li2000,Kobayashi&Zhang2003b} and after which the shocked region begins to decelerate in a self-similar manner \citep{Blandford&McKee1976}; this occurs on a timescale in the observer frame which is roughly commensurate with the active duration of the relativistic jet (i.e. $\lesssim 1$ day in the case of AT 2022cmc if its X-ray emission is indicative).  The peak of the early optical phase, which indeed conceivably coincides with the peak of the jetted X-ray phase, may thus be reasonably interpreted as the onset of the FS self-similar deceleration phase, regardless of whether the optical emission originates from the FS or RS.

This deceleration radius is roughly given by
\begin{align}
R_{\rm dec}&\simeq2c\Gamma_0^2t_{\rm dec}\simeq6.5\times10^{16}{\,\rm cm\,}\left(\frac{\Gamma_{0}}{5}\right)^{2}\left(\frac{t_{\rm dec}}{0.5{\,\rm day}}\right)\ ,
\label{eq:Rdec}
\end{align}
where $c$ is the speed of light, $t_{\rm dec}$ is the deceleration time measured in the engine's rest frame, and $\Gamma_0$ is the Lorentz factor of the emitting region at the deceleration radius.  We have normalized $\Gamma_0$ to a value $=5$, motivated by an extrapolation of the values obtained by the equipartition analysis (Fig.~\ref{fig:equipartition}) to the optical peak time in the narrow jet scenario ($\theta_{\rm j}=0.1$). Using also the inferred external density profile (Eq.~\ref{eq:profile}) we are thus able to estimate the total jet energy:
\begin{align}
E_{\rm j,iso}&=\frac{4\pi}{3-k}m_{\rm p} c^2 \Gamma_0^2R_{\rm dec}^3 n(R_{\rm dec})
    \label{eq:E_jiso}\\
&\simeq9.1\times10^{52}{\,\rm erg\,}\left(\frac{\Gamma_0}{5}\right)^{4}\left(\frac{t_{\rm dec}}{0.5\,\rm day}\right)\left(\frac{n_{17}}{300\,\rm cm^{-3}}\right)\ ,
    \nonumber
\end{align}
where in the second line we have taken $k=2$.  This is comparable to the minimum radiated X-ray energy $E_{\rm X,iso}\simeq1.3\times10^{53}\,\rm erg$, though our model presented in the next section will require larger energies $E_{\rm j,iso}>E_{\rm X,iso}$. The beaming-corrected jet energy,
\begin{align}
E_{\rm j}=\frac{\theta_{\rm j}^2}{2}E_{\rm j,iso}\simeq5.0\times10^{50}{\,\rm erg\,}\theta_{\rm j,-1}^2E_{\rm j,iso,53}, .
    \label{eq:Ej}
\end{align}
is considerably smaller.  Here and hereafter we employ the convention $Q_x = Q/10^x$ for quantities in cgs units except for the density profile normalization $n_{17}$ in Eq.~\eqref{eq:profile}.

The net jet energy in Eq.~\eqref{eq:Ej} is roughly 10 times larger than the equipartition energy $E_{\rm eq}$ found in the previous section, consistent with the latter representing a lower limit on the total jet energy (also not including an potential hot proton component).  A comparison between $E_{\rm j}$ and $E_{\rm eq}$ can be used to constrain the standard microphysical shock parameters $\epe$ and $\epb$, which represent the ratios of the energies placed into non-thermal electrons and magnetic fields, respectively, relative to the post-shock thermal energy. Following \citet{BarniolDuran+2013}, for the narrow-jet case we have
\begin{align}
E_{\rm tot}\simeq(1+\epe^{-1})^{3/5}\left[\frac{11}{17}\epsilon^{-2/5}+\frac{6}{17}\epsilon^{3/5}\right]E_{\rm eq}\ ,
\end{align}
where $\epsilon\equiv(11/6)(\epb/\epe)$.  Equating $E_{\rm tot}$ with $E_{\rm j}$, we thus find
\begin{align}
\epb\sim10^{-3}\,\varepsilon_{\rm e,-1}^{-1/2}\left(\frac{E_{\rm eq}/E_{\rm j}}{0.1}\right)^{5/2}\, ,
    \label{eq:epb}
\end{align} 
in reasonable accord with the (admittedly loose) constraints on $\epb$ and $\epe$ found in the context of gamma-ray burst (GRB) afterglow modeling (e.g., \citealt{Ryan+2015}).

For the wide spreading-jet scenario ($\theta_{\rm j}=1/\Gamma$), the isotropic jet energy becomes $E_{\rm j,iso}\simeq5.7\times10^{51}{\,\rm erg\,}(\Gamma_0/2.5)^4(n_{17}/300\,\rm cm^{-3})$, where the initial Lorentz factor $\Gamma_0\simeq2.5$ is obtained by extrapolating the equipartition results to early times in the same way as the narrow jet scenario.  Insofar that this implies an energy much smaller than the radiated X-ray energy, this disfavors the wide-jet scenario on grounds that it would require an extremely high X-ray radiative efficiency.  We note that Eq.~\eqref{eq:E_jiso} can be used to estimate the total energy even if the jet is laterally expanding at $\gtrsim5\,\rm days$ unless the jet was already spreading at the initial stage. 

Additional information comes from the early optical emission, if interpreted as synchrotron emission from the jet afterglow \citep[e.g.,][]{Sari+1998}.  By fitting a power-law spectrum $F_{\nu}\propto \nu^{\beta}$ to the early optical emission, \citet{Andreoni+2022} found a best-fit spectral index $\beta = -1.32\pm0.18$.  Assuming slow-cooling electrons, this implies a power-law distribution $dN/dE \propto E^{-p}$ of non-thermal electron energy with an index $p=3.64\pm0.36$, much steeper than predicted for Fermi acceleration at ultrarelativistic shocks, $p=2-2.2$ \citep{Bell1978,Blandford&Ostriker1978,Keshet&Waxman2005} or found by GRB afterglow modeling, $p\simeq2-3$ \citep[e.g.,][]{Fong+2015}.  In contrast, assuming the optical emission to be in the fast cooling regime gives a more reasonable value, $p=2.64 \pm 0.36$, consistent with the theoretical prediction and observations. Therefore, we infer the optical emission likely arises from fast-cooling electrons.

 Emission from fast cooling electrons disfavors the RS as the origin of the optical emission.  The RS light curve peaks when the RS has completely crossed the unshocked-jet material because at this point the number of emitting particles reaches a maximum. After the crossing phase, no additional particles are accelerated, resulting in a sharp cut-off in the spectrum above the cooling frequency.  As with the radio/mm emission, the observed power-law decay of the optical light curve after peaking thus favors the FS instead of the RS as the origin of the optical emission.  We note that the RS and hence the particle injection may be still present after the optical peak as suggested by the long-lived X-rays. However, if the energy injection rate is proportional to the X-ray luminosity, the predicted decay of the RS luminosity $L_{\rm RS}\propto L_{\rm X}\propto t^{-2}$ is steeper than the observed optical decay, unless some additional mechanism acts to offset this decline.  

In the fast cooling regime the synchrotron luminosity from the decelerating FS is furthermore independent of the ambient density \citep[e.g.,][]{Granot&Sari2002}:
\begin{align}
\nu L_\nu&\stackrel{p=2.6}{\simeq}9.1\times10^{45}{\,\rm erg\,s^{-1}\,}{\varepsilon}_{\rm e,-1}^{p-1}{\varepsilon}_{\rm B,-3}^{\frac{p-2}{4}}E_{\rm j,iso,53}^{\frac{p+2}{4}}\theta_{\rm j,-1}^2
    \nonumber\\
&\left(\frac{\Gamma_0}{5}\right)^2\left(\frac{t}{\rm day}\right)^{\frac{2-3p}{4}}\left(\frac{1+z}{2.19}\right)^{\frac{4-p}{2}}\left(\frac{\nu}{5\times10^{14}\,\rm Hz}\right)^{-\frac{p}{2}}\ ,
    \label{eq:Lopt}
\end{align}
where the numerical values are calculated for $p = 2.6$ and have taken into account the suppression factor $(\Gamma\theta_{\rm j})^2$ given the angular size of the emitting region $\pi \theta_{\rm j}^{2}$ for $\theta_{\rm j}<1/\Gamma$.  The agreement between these predictions and the observed optical flux provides a consistency check on the FS model.

\subsection{Light Curve Calculation}\label{sec:lc}

\begin{table}
\begin{center}
\caption{Model parameters for the synchrotron light curve.}
\label{table:parameter}
\begin{tabular}{lc}
\hline
Parameter&Value\\
\hline
$n_{17}$: Density of external medium at $10^{17}$ cm&$200\,\rm cm^{-3}$\\
$k$: Power-law slope of radial density profile&1.8\\
$\theta_{\rm j}$: Jet half-opening angle&0.15\\
$E_{\rm j,iso}$: Isotropic jet energy&$4\times10^{53}$ erg\\
$\Gamma_0$: Initial Lorentz factor of shocked gas&5\\
$p$: Slope of the electron energy distribution&2.9\\
$\epe$: Energy fraction of non-thermal electrons&0.2\\
$\epb$: Energy fraction of magnetic field&0.002\\
\hline
\end{tabular}
\end{center}
\end{table}

\begin{figure}
\begin{center}
\includegraphics[width=85mm, angle=0]{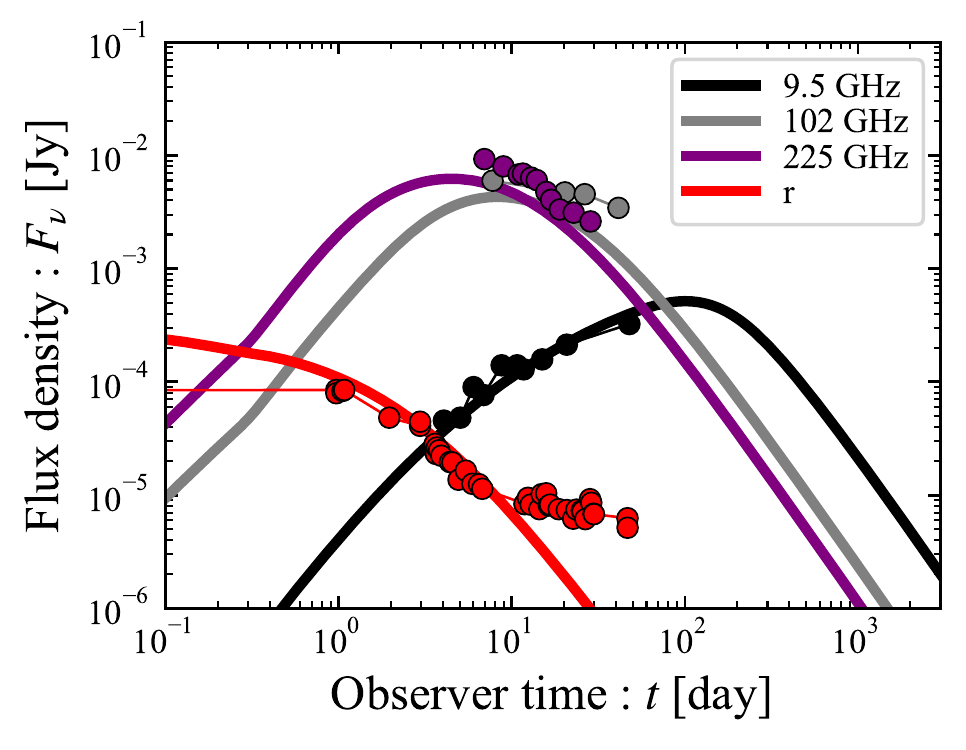}
\caption{Synchrotron light curve model (lines) fit to the optical/radio data (circles) for AT 2022cmc, considering only emission from the forward shock. The late-time r-band ($\gtrsim5\,\rm days$) likely arises from a separate thermal emission component unrelated to the FS, similar to that observed in Swift J2058+05 and other optically-selected TDEs. The parameters of the model are given in Table.~\ref{table:parameter}.}
\label{fig:lc}
\end{center}
\end{figure}

Guided by the preliminary considerations in the previous section, we model the light curve of AT 2022cmc assuming the radio and early optical emission both originate from the decelerating FS. The synchrotron light curve is calculated in the same manner as outlined in \cite{Bruni+2021,Ricci+2021}, but we adopt the prescription of \cite{Granot&Sari2002} to smooth the spectrum and introduce the suppression factor $(\Gamma\theta_{\rm j})^2$ on the flux to account for the finite emitting size of the jet, as mentioned above. The parameters of the model are summarized in Table~\ref{table:parameter}, with several of their values already motivated by the analysis in the previous section.

\begin{figure}
\begin{center}
\includegraphics[width=85mm, angle=0]{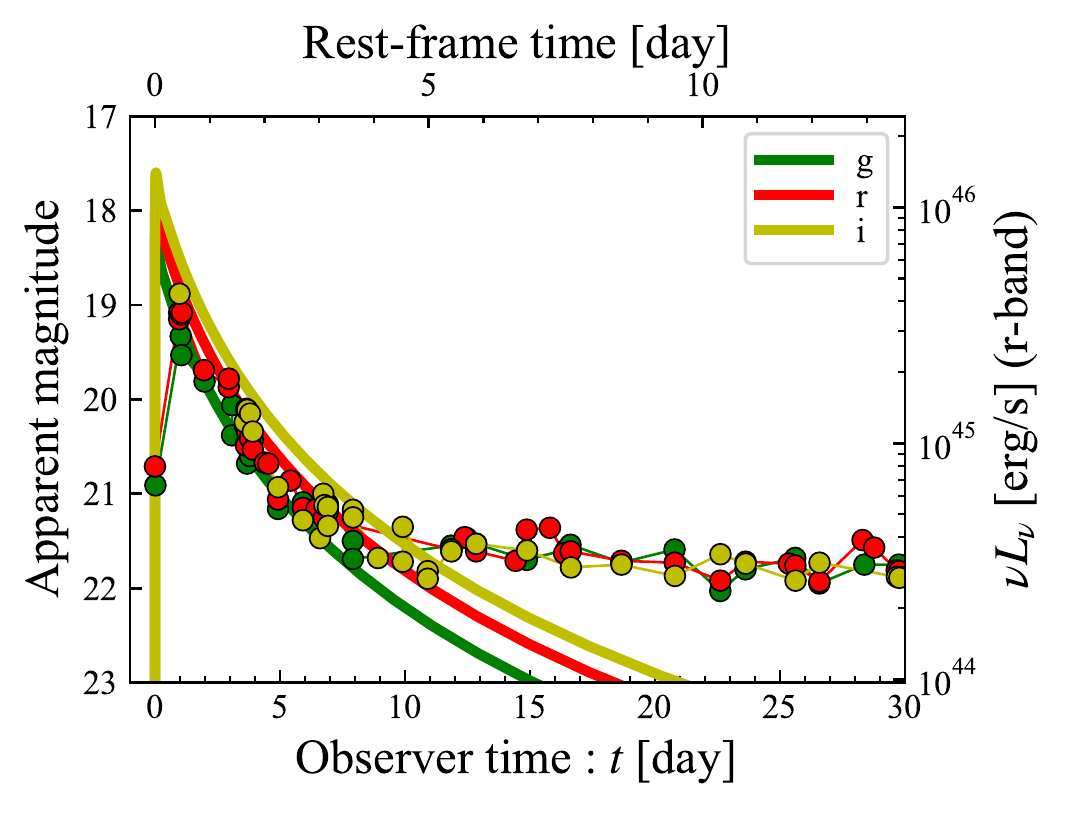}
\caption{Optical light curves for AT 2022cmc compared to our synchrotron afterglow model. The right vertical axis denotes the luminosity for r-band. The luminosities at g and i-bands are roughly 1.34 and 0.88 times larger than the r-band luminosity.}
\label{fig:lc2}
\end{center}
\end{figure}

Fig.~\ref{fig:lc} depicts a light curve model that reasonably reproduces the radio and optical data, which we show for comparison with circles.  The adopted parameter values of this model (Table~\ref{table:parameter}) were found heuristically by exploring values around those hinted by the equipartition analysis, rather than through a systematic parameter scan.  Fig.~\ref{fig:lc2} shows just the optical data, now broken down into separate colors.  As already mentioned, the light curve is comprised of two parts: an early red peak followed by blue plateau \citep{Andreoni+2022,Pasham+2022}. Insofar as the late plateau ($\gtrsim5\,\rm days$) is better described as thermal emission of temperature $\simeq(2-4)\times10^4\,\rm K$ similar to optical TDEs \citep{VanVelzen+2021,hammerstein+2023}, we ignore this component and focus on fitting just the early peak phase.

\begin{figure}
\begin{center}
\includegraphics[width=85mm, angle=0]{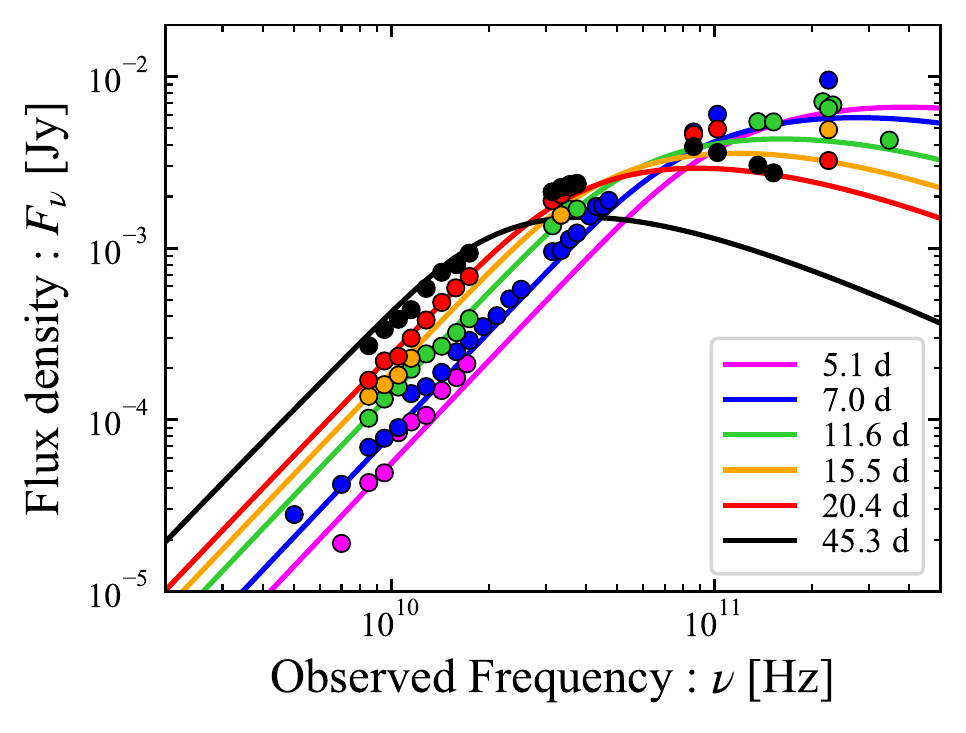}
\caption{Radio spectrum of AT 2022cmc at each epoch (times measured in the observer frame; from \citealt{Andreoni+2022}) compared to our model predictions.}
\label{fig:spectrum}
\end{center}
\end{figure}

\begin{figure}
\begin{center}
\includegraphics[width=85mm, angle=0]{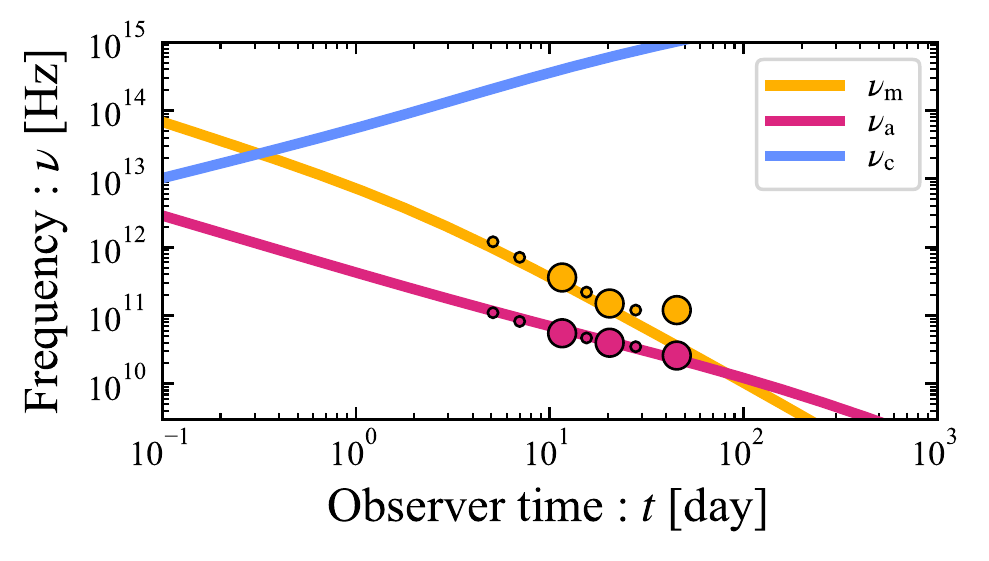}
\caption{Time evolution of the characteristic synchrotron ($\num$), SSA ($\nua$), and cooling ($\nuc$) frequencies based on fitting the observed radio/mm spectrum, compared to the model predictions.}
\label{fig:nu}
\end{center}
\end{figure}

Figs.~\ref{fig:spectrum} and \ref{fig:nu} show, respectively, the radio spectrum of the model and the data at each epoch and the time evolution of key synchrotron break frequencies.  Although our favored model largely agrees with the observations (within a factor of a few) at most epochs, there is a noticeable discrepancy in the late-time spectrum near day 45.3.  Around $\sim100\,\rm GHz$, our theoretical spectrum underestimates the observed flux by a factor of $3-4$. We speculate that this excess could reflect additional contributions to the observed emission from different angular portions of the jet FS not captured by our one-zone model (e.g., the ``sheath'' surrounding the jet core; \citealt{Mimica+2015,Generozov+2017}) or energy injection from slower material gradually catching up with the FS region \citep{Berger+2012}, as was previously proposed to explain the late-time brightening in Sw J1644+57.  Future observations will test this possibility.

Combining the isotropic jet energy $E_{\rm j,iso}$ of our reasonably good-fitting model with the measured X-ray energy $E_{\rm X,iso}$, we can constrain the prompt radiative efficiency of the jet,
\begin{align}
\varepsilon_{\rm X}\equiv\frac{E_{\rm X,iso}}{E_{\rm j,iso}+E_{\rm X,iso}}\gtrsim0.2\ ,
\end{align}
where we have used $E_{\rm j,iso}=4\times10^{53}\,\rm erg$ and $E_{\rm X,iso}=10^{53}\,\rm erg$. The latter (and hence also $\varepsilon_{\rm X}$) is a lower-limit because the X-ray luminosity $L_{\rm X} \propto t^{-2}$ had already begun decaying by the time observations commenced at $t \simeq5\,\rm days$. The extrapolation of this luminosity to $t\simeq1\,\rm day$ results in the X-ray energy of $E_{\rm X,iso}\simeq5\times10^{53}\,\rm erg$ and relatively high emission efficiency of $\varepsilon_{\rm X}\simeq0.6$, comparable to that found for Swift J1644+57, $\varepsilon_{\rm X}\simeq0.5$ \citep{Berger+2012} and the gamma-ray efficiencies of GRBs ($\varepsilon_{\gamma}\simeq0.4-0.6$, \citealt{Fong+2015}, but see also \citealt{Beniamini+2016}).

\section{Discussion}\label{sec:discussion}

\subsection{Early Reverse Shock Emission}
\label{sec:RS}

While our analysis favors the FS instead of the RS as the origin of the radio/mm and early optical emission from AT2022cmc, the RS may still generate observable emission.  This will occur on timescales $\sim 1$ day, comparable to the duration of the X-ray/optical peak, when the RS is crossing the bulk of the jet material and its emission reaches a maximum \citep{Sari&Piran1999b,Kobayashi2000}.  The RS emission offers a potential probe of the unshocked jet material \citep{Giannios&Metzger2011,Metzger+2012}, particularly its bulk Lorentz factor and magnetization.  In this section we discuss the detectability of the RS emission and the implications of its presence or absence in AT 2022cmc and similar future events.

During the RS crossing phase, the Lorentz factor of the post-shock gas $\Gamma_{\rm sh}$ and the Lorentz factor of the RS relative to the unshocked jet $\tilde{\Gamma}_{\rm RS}$, are determined by two parameters \citep{Sari&Piran1995}. One is the Lorentz factor of the unshocked jet, $\Gamma_{\rm j}$, and the other is the density ratio of the unshocked-jet material to the external medium:
\begin{align}
f\equiv\frac{n_{\rm j}}{n_{\rm ext}}\simeq5.9\,L_{\rm j,iso,48}\Gamma_{\rm j,1}^{-2}\left(\frac{n_{17}}{300\,\rm cm^{-3}}\right)^{-1}\ ,
\end{align}
where the comoving density of the unshocked jet with kinetic power $L_{\rm j,iso}$ is given by
\begin{align}
n_{\rm j}=\frac{L_{\rm j,iso}}{4\pi R^2 m_{\rm p} c^3\Gamma_{\rm j}^2}\simeq1800{\,\rm cm^{-3}\,}L_{\rm j,iso,48}\Gamma_{\rm j,1}^{-2}R_{17}^{-2}\ ,
\end{align}
and we have adopted the external density profile (Eq.~\ref{eq:profile}) with $k=2$, making $f$ conveniently independent of radius.

For $\Gamma_{\rm j}^2\gg f$, the RS is relativistic and the Lorentz factors of the RS and shocked material are given by
\begin{align}
\tilde{\Gamma}_{\rm RS}&\simeq\left(\frac{\Gamma_{\rm j}^2}{4f}\right)^{1/4}\simeq1.4\,\Gamma_{\rm j,1}L_{\rm j,iso,48}^{-1/4}\left(\frac{n_{17}}{300\,\rm cm^{-3}}\right)^{1/4}\ ,
\\
\Gamma_{\rm sh}&\simeq\Gamma_{\rm j}\left(\frac{f}{4\Gamma_{\rm j}^2}\right)^{1/4}\simeq3.5\,L_{\rm j,iso,48}^{1/4}\left(\frac{n_{17}}{300\,\rm cm^{-3}}\right)^{-1/4}\ .
    \label{eq:Gam_sh}
\end{align}
Note that $\Gamma_{\rm sh}$ is independent of $\Gamma_{\rm j}$, and roughly coincidences with initial Lorentz factor of the FS-emitting material defined earlier, $\Gamma_0$. In particular, the fiducial value of our FS afterglow model $\Gamma_0=5$ (Table \ref{table:parameter}) is obtained for $L_{\rm j,iso}\simeq2.8\times10^{48}{\,\rm erg\,s^{-1}\,}(n_{17}/200\,\rm cm^{-3})$, consistent with $L_{\rm j,iso} \simeq E_{\rm j,iso}/t_{\rm j}$ given the total jet energy $E_{\rm j,iso} \sim 10^{53}\,\rm erg$ and peak jet duration $t_{\rm j} \sim 1$ day.

\begin{figure}
\begin{center}
\includegraphics[width=85mm, angle=0]{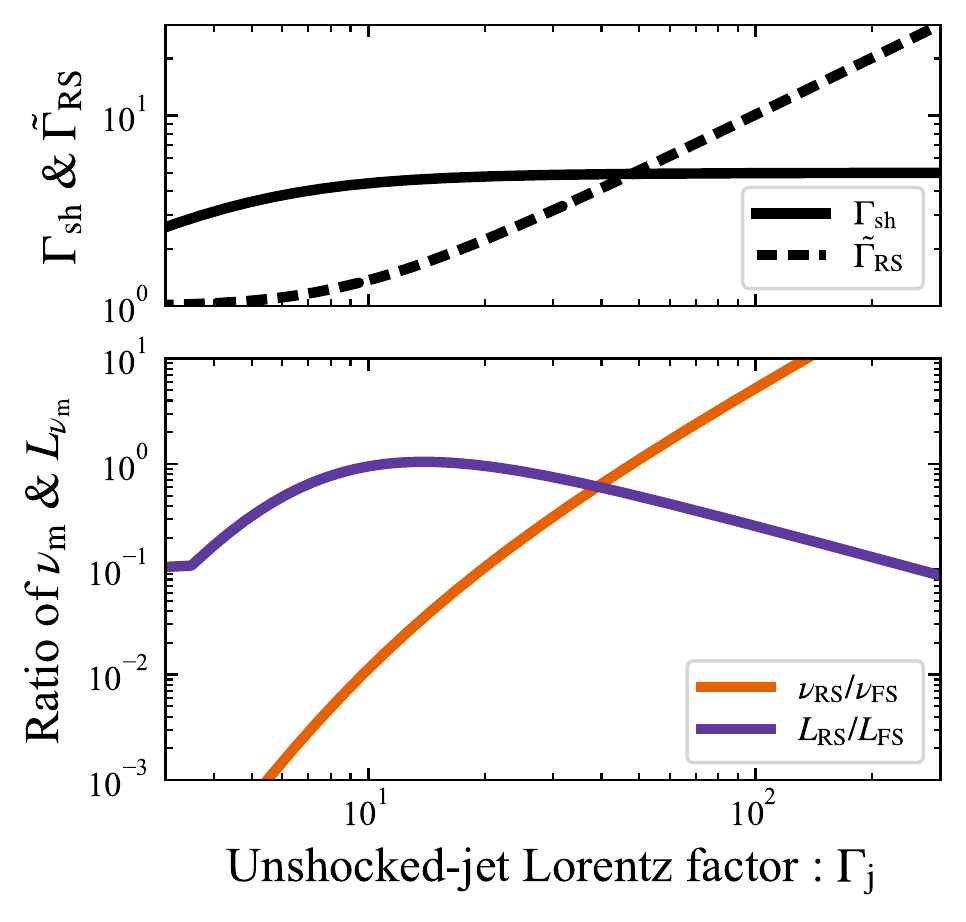}
\caption{({\bf Top}) Lorentz factors of the shocked jet material ($\Gamma_{\rm sh}$, solid) and the RS ($\tilde{\Gamma}_{\rm RS}$, dashed) as a function of the unshocked-jet Lorentz factor $\Gamma_{\rm j}$. The parameters are set as $L_{\rm j,iso}=2.8\times10^{48}{\,\rm erg\,s^{-1}\,}(n_{17}/200\,\rm cm^{-3})$ so that $\Gamma_{\rm sh}=5$ for $\Gamma_{\rm j}\gg10$ (see Eq.~\ref{eq:Gam_sh}). ({\bf Bottom}) Ratios of the characteristic synchrotron frequency ($\num$, orange) and luminosity ($L_{\num}$, purple) of the RS to FS.}
\label{fig:shock}
\end{center}
\end{figure}

The top panel of Fig.~\ref{fig:shock} depicts both $\Gamma_{\rm sh}$ and $\tilde{\Gamma}_{\rm RS}$ as a function of $\Gamma_{\rm j}$ obtained by solving the exact shock jump condition in \cite{Sari&Piran1995}. To obtain the shocked Lorentz factor of 5 indicated by our FS afterglow analysis, the Lorentz factor of the unshocked jet is constrained to obey $\Gamma_{\rm j}\gtrsim10$, similar to inferred for Swift J1644+57 (e.g., \citealt{Metzger+2012}).  While $\Gamma_{\rm sh}$ is independent of $\Gamma_{\rm j}$, the RS Lorentz factor increases with $\Gamma_{\rm j}$, rendering the RS emission sensitive to its value.  In particular, the RS becomes relativistic for
\begin{align}
\Gamma_{\rm j}\gtrsim\Gamma_{\rm j,cr}\simeq14\,L_{\rm j,iso,48}^{1/4}\left(\frac{n_{17}}{300\,\rm cm^{-3}}\right)^{-1/4}\ .
\end{align}

We now estimate the SED of the RS emission from AT 2022cmc at its peak, which as already mentioned is expected to coincide with the early optical peak at $t \sim 1$ day.  This is done by considering the ratio between the characteristic synchrotron frequency and luminosity of the FS versus the RS. Assuming the same microphysical parameter $\varepsilon_{\rm B}$ for both the FS and RS, the magnetic field strength of the post-shock gas will also be identical due to the pressure equilibrium across the contact discontinuity; the bulk Lorentz factor of the emitting gas $\Gamma_{\rm sh}$ behind both shocks is also the same.  However, the random Lorentz factors of the electrons behind the FS and RS differ because of their different respective shock strengths (shock Lorentz factors).  Taking these points into account, the ratio of the characteristic frequencies of the FS to RS emission can be estimated as (e.g., \citealt{Giannios&Metzger2011})
\begin{align}
\frac{\nu_{\rm m,RS}}{\nu_{\rm m,FS}}=\frac{(\tilde{\Gamma}_{\rm RS}-1)^2}{(\Gamma_{\rm sh}-1)^2}\simeq0.17\,\Gamma_{\rm j,1}^{2}L_{\rm j,iso,48}^{-1}\left(\frac{n_{17}}{300\,\rm cm^{-3}}\right)\ ,
    \label{eq:nu_ratio}
\end{align}
where the last equality holds only for $\Gamma_{\rm j}>\Gamma_{\rm j,cr}$. The bottom panel of Fig.~\ref{fig:shock} depicts the behavior of this ratio as a function of $\Gamma_{\rm j}$. 

Since the synchrotron emissivity is the same for both the FS and RS emission regions, and assuming the electrons behind both shocks are in a slow-cooling regime, the ratio of their luminosities at $\num$ can be obtained by considering the number of accelerated particles at each. At the peak of the FS, the number of particle in the post FS region is given by the particles swept up by the FS, $\sim E_{\rm j,iso}/(m_{\rm p}c^2\Gamma_{\rm sh}^2)$. For the RS, the number is given by the particles within the shell $\sim E_{\rm j,iso}/(m_{\rm p}c^2\Gamma_{\rm j})$. Therefore, the luminosity ratio is given by
\begin{align}
\frac{L_{\rm \num, RS}}{L_{\rm \num, FS}}\simeq\frac{\Gamma_{\rm sh}^2}{\Gamma_{\rm j}}\simeq1.2\,\Gamma_{\rm j,1}^{-1}L_{\rm j,iso,48}^{1/2}\left(\frac{n_{17}}{300\,\rm cm^{-3}}\right)^{-1/2}\ .
    \label{eq:L_ratio}
\end{align}
Again, the last equality holds for $\Gamma_{\rm j}>\Gamma_{\rm j,cr}$. We show this ratio as a function of $\Gamma_{\rm j}$ in the bottom panel of Fig.~\ref{fig:shock}. It should be noted that when the shock becomes Newtonian and its velocity becomes smaller than a critical value $v_{\rm DN}\simeq0.2\,c\,\varepsilon_{\rm e,-1}^{-1/2}$ for $p=2.6$, the minimal random Lorentz factor becomes a few and the so-called deep Newtonian regime sets in \citep{Huang&Cheng2003,Sironi&Giannios2013}. In this regime, the number of the accelerated particles is suppressed relative to the standard case by a factor of $\simeq(v/v_{\rm DN})^2$, where $v$ is the shock velocity.  This effect is included in our calculation and is responsible for the break in the luminosity ratio at $\Gamma_{\rm j}\simeq3$ in Fig.~\ref{fig:shock}.

\begin{figure}
\begin{center}
\includegraphics[width=85mm, angle=0]{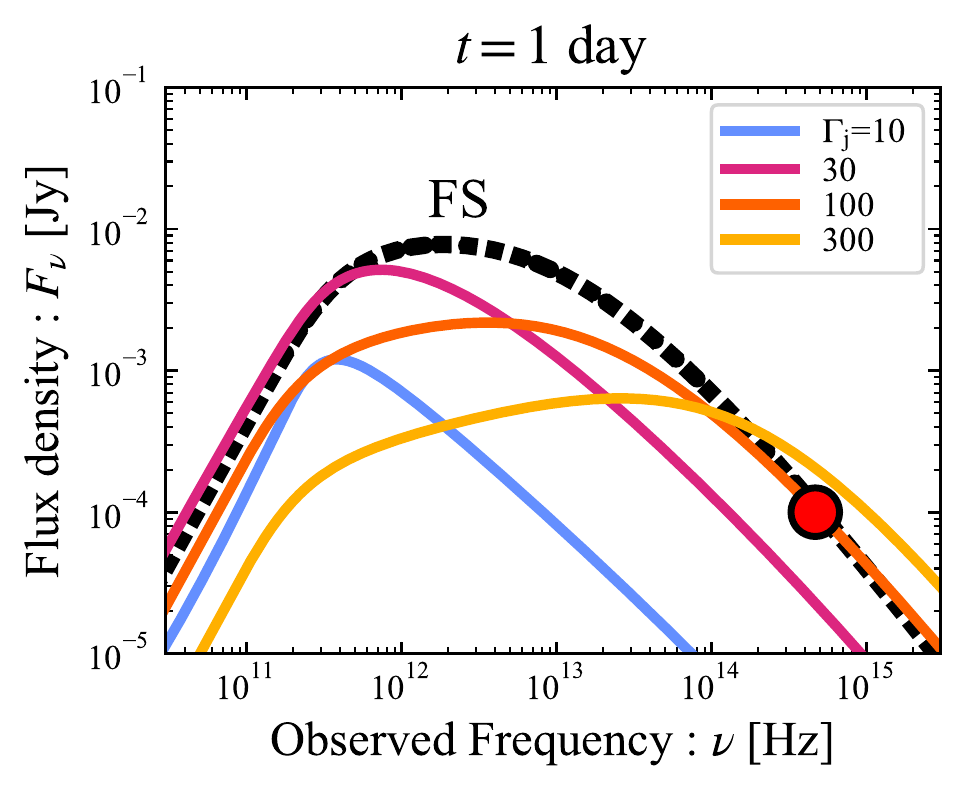}
\caption{The estimated SED of the RS emission from AT 2022cmc at $1\,\rm day$, when it reaches peak luminosity around the time of the optical maximum. The red point denotes the observed r-band flux at this epoch.}
\label{fig:spectrum2}
\end{center}
\end{figure}

We estimate the SED of the RS near the time of the optical peak in Fig.~\ref{fig:spectrum2}. The characteristic frequency $\num$ and its luminosity $L_{\num}$ are obtained by scaling those of the FS according to Eqs.~\eqref{eq:nu_ratio} and \eqref{eq:L_ratio}. We account for suppression of the flux due to SSA at the RS as well as the FS.  We find for sufficiently large jet Lorentz factors $\Gamma_{\rm j}\gtrsim30$, SSA at the RS does not contribute to shaping the spectrum as much as the FS, due to the lower density and higher temperature of the former. For even larger jet Lorentz factor $\Gamma_{\rm j}\gtrsim100$, the RS may in fact dominate the optical emission, contrary to our earlier inference that the optical emission in AT 2022cmc is likely to be dominated by the FS (\S \ref{sec:preliminary}).  In the fast cooling regime, the luminosity is estimated by using Eq.~\eqref{eq:Lopt}, which holds after the beginning of the deceleration, and Eq.~\eqref{eq:Gam_sh} for $\Gamma_0$ and scaling relations of \eqref{eq:nu_ratio} and \eqref{eq:L_ratio}, \begin{align}
\left(\nu L_{\nu}\right)_{\rm RS}\stackrel{p=2.6}{\simeq}&2.6\times10^{45}{\,\rm erg\,s^{-1}\,}{\varepsilon}_{\rm e,-1}^{p-1}{\varepsilon}_{\rm B,-3}^{\frac{p-2}{4}}E_{\rm j,iso,53}^{\frac{p+2}{4}}\theta_{\rm j,-1}^2
    \nonumber\\
&\Gamma_{\rm j,1}^{p-2}L_{\rm j,iso,48}^{\frac{3-p}{2}}\left(\frac{n_{17}}{300\,\rm cm^{-3}}\right)^{\frac{p-3}{2}}\ ,\label{eq:Lopt_RS}
\end{align}
where we have omitted the dependence on $t$, $z$, and $\nu$, which are the same as for Eq.~\eqref{eq:Lopt}. Note this equation is valid for $\nu_{\rm m,RS}<\nuc$ and $\Gamma_{\rm j}>\Gamma_{\rm j,cr}$.
The non-detection of an extra component in the optical light curve thus places a weak upper limit on the unshocked jet Lorentz factor for our fiducial parameters in Table~\ref{table:parameter}:
\begin{align}
&\Gamma_{\rm j}\lesssim100\,\left(\frac{{\varepsilon}_{\rm e}}{0.2}\right)^{-\frac{p-1}{p-2}}\left(\frac{{\varepsilon}_{\rm B}}{0.002}\right)^{-\frac{1}{4}}\left(\frac{E_{\rm j,iso}}{4\times10^{53}{\,\rm erg}}\right)^{-\frac{p+2}{4(p-2)}}
    \nonumber\\
&\left(\frac{\theta_{\rm j}}{0.15}\right)^{-\frac{2}{p-2}}\left(\frac{L_{\rm j,iso}}{2.8\times10^{48}{\,\rm erg\,s^{-1}}}\right)^{\frac{p-3}{2(p-2)}}\left(\frac{n_{17}}{200\,\rm cm^{-3}}\right)^{-\frac{p-3}{2(p-2)}}\ .
\end{align}

Finally, we note that our analysis has assumed a jet that is relatively weakly magnetized, by the deceleration radius $\sim 10^{17}$ cm (Eq.~\ref{eq:Rdec}).  In the jet were instead to remain highly magnetized, the jet energy is still transferred from the FS and its emission remains largely unchanged; however,  the strength of the RS and hence its luminosity can be strongly suppressed relative to our estimates here for a highly magnetized jet (e.g., \citealt{Zhang&Kobayashi2005,Giannios+2008}).

\subsection{The Origin of AT 2022cmc}

Although AT 2022cmc was interpreted as a jetted TDE \citep{Andreoni+2022,Pasham+2022}, there are reasons to be cautious about this connection.  One is that the location of the transient within its host galaxy is not yet resolved, making a nuclear origin impossible to confirm, at least until the transient has faded.  Secondly, the inferred duration of the jetted activity, as implied by the length of the early optical and X-ray emission phase, is over an order of magnitude shorter than the typical fallback timescale $\gtrsim$ weeks of the stellar debris in a TDE for a typical SMBH mass $\sim 10^{6}M_{\odot}$.  This timescale problem could be alleviated by invoking a much lower-mass (e.g. intermediate mass) black hole, or by considering the tidal disruption of a white dwarf instead of a main-sequence star (e.g., \citealt{Krolik&Piran2011}); however, these scenarios are not obviously consistent with the $\gtrsim$ months-long duration of the thermal optical emission assuming the latter is also powered by accretion onto the SMBH (see below).   

Nevertheless, the luminous blue optical plateau phase, whose colors and estimated bolometric luminosity of $\sim10^{45}\,\rm erg\,s^{-1}$ are indeed similar to those observed in optically-selected TDEs (e.g.~\citealt{VanVelzen+2021,hammerstein+2023}) as well as the jetted TDE candidate Swift J2058+05 \citep{Cenko+2012}, but not consistent with most other transient classes.  Such high luminosity events are still rare among the optically-selected TDE population and are indeed characterized by spectra lacking emission or absorption lines \citep{hammerstein+2023}, qualitatively similar to the featureless spectrum of the plateau phase in AT 2022cmc \citep{Andreoni+2022}.  Extremely luminous TDEs are challenging to explain in models for TDE optical emission which invoke shocks between tidal debris streams (e.g., \citealt{Piran+2015,Ryu+2020})\footnote{Following analytic estimates presented in \cite{Ryu+2020}, we find that reproducing a peak luminosity $\sim10^{45}\,\rm erg\,s^{-1}$ would require the disruption of an exceptionally massive $m_\star\gtrsim100\,\Msun$ star.} and the specific implication in AT 2022cmc that efficient accretion onto the SMBH has already begun for the jet to be launched.\footnote{It should be noted that these bright featureless TDEs with radiated energy of $\sim10^{52}\,\rm erg$ are free from the inverse energy crisis advocated for optical TDEs \citep{Piran+2015,Svirski+2017}.} The large luminosity is also in tension with reprocessing wind models (e.g., \citealt{Metzger&Stone2016,Dai+2018,Lu&Bonnerot2020}) due to the large required ejecta mass \citep{Matsumoto&Piran2021}.  

In the ``cooling envelope'' emission model (\citealt{Metzger2022b}; see also \citealt{Coughlin&Begelman2014}), TDE thermal optical emission originates from a quasi-spherical hydrostatic envelope radiating near the SMBH Eddington limit, which to explain the observed optical plateau luminosity $\simeq 10^{45}$ erg s$^{-1}$ of AT 2022cmc would require a SMBH mass $M_{\rm BH} \sim 10^{7}M_{\odot}$.  To reproduce the long duration of the optical phase $\gtrsim$ months if interpreted as the Kelvin-Helmholtz cooling time of the envelope, would require the disruption of a massive star $m_{\star} \gtrsim 10M_{\odot}$.  Alternatively, a plateau light curve phase is generated if the SMBH accretion flow continually resupplies the envelope with energy in a regulated manner; however, the required long-lived accreting SMBH engine is again challenged by the much shorter jet activity period suggested by the X-ray emission.

An alternative model for AT 2022cmc is the core collapse of a massive star that gives birth to an energetic central compact object, such as a millisecond magnetar (e.g., \citealt{Thompson+2004}) or accreting black hole (e.g., \citealt{Quataert&Kasen2012,Dexter&Kasen2013,Perna+18}).  A magnetar engine rotating near its break-up velocity (spin period $P \sim 1$ ms) with a long energy injection (e.g., dipole spin-down) timescale $t_{\rm sd} \sim 1$ day, was proposed by \citet{Metzger+2015d} to explain Swift J2058+05 and the extremely luminous transient ASASSN-15lh (\citealt{Dong+2016}; also claimed to be a TDE by \citealt{Leloudas+2016}); a portion of the magnetar's spin-down energy goes into a bipolar relativistic jet of active duration $t_{\rm sd}$ which is capable of breaking through the expanding stellar envelope (e.g., \citealt{Margalit+2018}) while the remainder is thermalized behind the ejecta, powering extremely luminous SN emission (e.g., \citealt{Kasen&Bildsten2010,Woosley2010,Vurm&Metzger21}).  A similar jetted-engine-boosted SN scenario was proposed by \citet{Metzger+2015d} to explain the over-luminous SN 2011kl associated with the ultra-long GRB 111209A \citep{Greiner+15}.

\begin{figure}
\begin{center}
\includegraphics[width=85mm, angle=0]{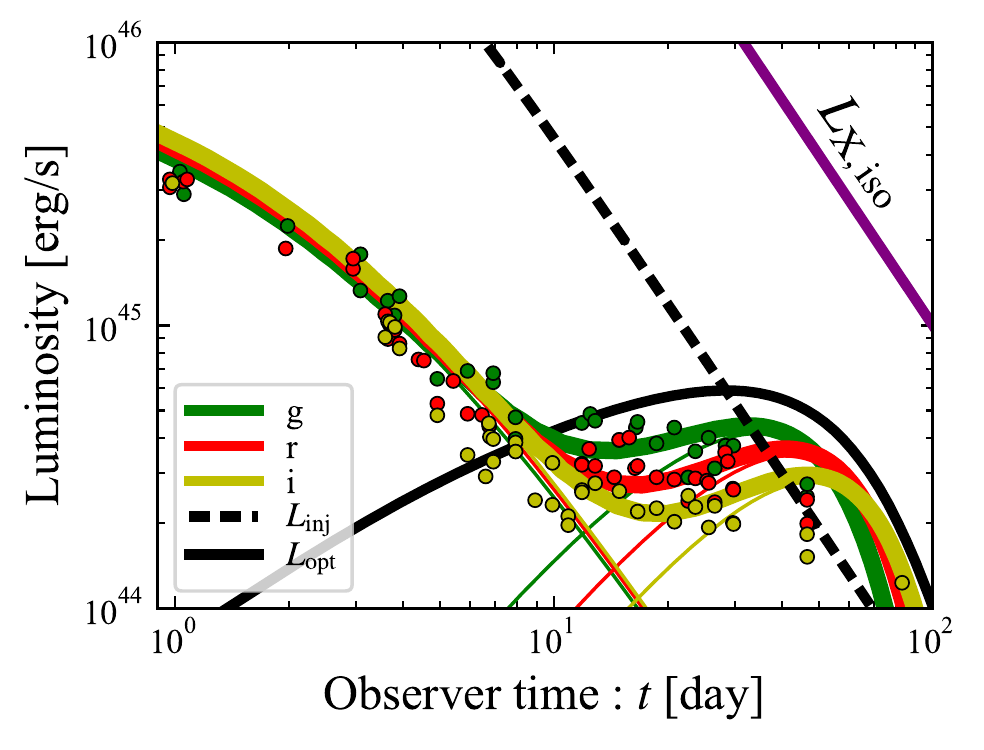}
\caption{Combined jet afterglow and engine-powered SN model fit to the optical light curves of AT2022 cmc.
The g, r, and i-band light curves are shown with thick green, red, and yellow curves, respectively, and are decomposed with thinner lines into separate contributions from the jet afterglow ($t \lesssim 10$ days; \S \ref{sec:lightcurve}) and SN ($t \gtrsim 10$ days).  For the engine-powered SN model \citep{Metzger+2015d} we assume an ejecta mass $3\,\Msun$, (initial) kinetic energy $10^{51}\,\rm erg$, and opacity $0.1\,\rm cm^{2}\,g^{-1}$.  Black solid and dashed curves show the bolometric optical/UV light curve of the SN and injected luminosity from the central engine, respectively. The latter is taken to be $\simeq3\,\%$ of a power-law fit to the observed X-ray luminosity $L_{\rm X,iso}=3\times10^{47}{\,\rm erg\,s^{-1}}(t/5\,\rm day)^{-2}$ (purple line; see text for discussion). We note that here we show the luminosity at each wavelength, which differs from the spectral luminosity shown in Fig.~\ref{fig:lc2}.} 
\label{fig:lc3}
\end{center}
\end{figure}

Fig.~\ref{fig:lc3} shows an example light curve model in the engine-powered SN scenario for AT 2022cmc. We adopt a simple one-zone model as \citet{Metzger+2015d} but with an energy injection rate to the ejecta from the central engine following $L_{\rm inj}=10^{47}/[1+(t/\rm day)^2]\,\rm erg\,s^{-1}$ (dashed line) motivated by the X-ray light curve (purple line). The ejecta mass and opacity are set to $M_{\rm ej}=3\,\Msun$ and $\kappa=0.1\,\rm cm^{2}\,g^{-1}$, respectively. The SN component can broadly reproduce the blue plateau up to $\sim100\,\rm days$, though color evolution resulting from the ejecta cooling is inevitable in a supernova model.  The initial kinetic energy is $E_{\rm kin,0}=10^{51}\,\rm erg$, but the result does not depend on this precise value as long as it is smaller than the injected energy $E_{\rm kin,0}<E_{\rm inj}\simeq10^{52}\,\rm erg$. 
Since the peak luminosity is roughly the same as the luminosity injected by the engine at the diffusion time $\sim\sqrt{\frac{3\kappa M_{\rm ej}}{4\pi cv_{\rm ej}}}$ \citep{Arnett1982}, the peak time of $\simeq30\,\rm days$ constrains the ejecta mass $M_{\rm ej}\simeq3\,\Msun\,E_{\rm inj,52}^{1/3}\kappa_{-1}^{-2/3}$. Given the adopted injection luminosity of $\simeq10^{47}\,\rm erg\,s^{-1}$ at 1 day and the extrapolated X-ray luminosity of $\simeq8\times10^{48}\,\rm erg\,s^{-1}$ at 1 day with the inferred efficiency of $\varepsilon_{\rm X}\simeq0.6$ and the beaming fraction of $\sim10^{-2}$, we find the central engine must share its energy roughly equally between its relativistic jet and that going into SN heating.  Such a partition is realized by dissipation of the striped wind in a millisecond magnetar scenario if the magnetic dipole is misaligned with the rotation axis by an angle $\simeq 0.4$ \citep{Margalit+2018}.  

We comment that such rapid magnetar birth spin periods $\sim 1$ ms are challenging to explain from binary star evolution models which include updated angular momentum transport schemes \citep{Fuller&Lu2022}.  Chemically homogeneous single-star models may fare better, though the necessarily high mass of such progenitors would likely require a larger SN ejecta mass $ \gtrsim 10M_{\odot}$ than our example model in Fig.~\ref{fig:lc3}.  On the other hand, our spherically-symmetric model could overestimate the light curve evolution timescale (and hence underestimate the ejecta mass) for a viewer along the rotation axis, where the expansion velocity is higher than the average. For an accreting black hole scenario, the long jet duration $\gtrsim5\,\rm days$ requires a long fallback timescale for the stellar envelope, thus requiring a giant stellar progenitor \citep[e.g.,][]{Quataert&Kasen2012}.

While the featureless spectra of AT 2022cmc taken during the plateau phase may pose a challenge to SN scenarios \citep{Andreoni+2022}, we note that both ASASSN-15lh \citep{Dong+2016} and SN 2011kl \citep{Greiner+15} were nearly featureless, which \citet{Mazzali+2016} found could be attributed to the Doppler broadening associated with high ejecta expansion velocities $\gtrsim 20,000$ km s$^{-1}$.  In our best fit model shown in Fig.~\ref{fig:lc3}, the large energy injected by the engine at early times prior to the optical peak $\sim 10^{52}$ erg when radiation is still trapped, indeed accelerates the ejecta to high velocities $\sim0.1\,c$, even if the initial explosion possessed a lower kinetic energy.

In principle the density of the external environment as probed by the FS can also provide clues as to the nature of the central engine.  In a TDE scenario, the inferred density profile $n \propto R^{-k}$ with $k \simeq 1.5-2$ (Fig.~\ref{fig:profile}) may be the result of gas created by star formation activity locally in the nuclear star cluster \citep{Generozov+2017} or a part of an accretion flow from larger radii onto the SMBH.  In particular, the measured density slope is close to the prediction $k=1.5$ for Bondi accretion onto the SMBH (e.g., \citealt{Quataert+1999}), in which case the density normalization $n_{17}\simeq300\,\rm cm^{-3}$ translates into a mass accretion rate $\dot{M}\sim10^{-3}\,\dot{M}_{\rm Edd}(M_{\rm BH}/10^{6}M_{\odot})^{-1/2}$, where $\dot{M}_{\rm Edd}$ is the Eddington accretion rate.  This would support the nucleus of AT 2022cmc being a low-luminosity AGN (e.g., \citealt{Ho1999}), which may not be surprising given the preference for TDEs to occur in post-starburst galaxies with likely recent AGN activity (e.g., \citealt{Arcavi+2014}) or the requirement of pre-existing magnetic flux in the galactic center environment from such a ``fossil'' disk in some TDE jet formation scenarios (e.g., \citealt{Tchekhovskoy+2014,Kelley+2014}).

On the other hand, in the SN scenario, the density slope $k=2$ could also be interpreted within in a massive star explosion scenario as that of a stellar wind, in which case the normalization $n_{17}\simeq300\,\rm cm^{-3}$ would correspond to that of stellar mass-loss rate of $\dot{M} \simeq10^{-4}\,\rm \Msun\,yr^{-1}$ for an assumed wind velocity of $1000\,\rm km\,s^{-1}$ ($A_*\sim10$ in the nomenclature of \citealt{Chevalier&Li1999}), similar to those which characterize the circumstellar environments of core-collapse SNe (e.g., \citealt{Chevalier&Fransson2006,Chevalier1998}; individual SNe are shown as gray crosses in Fig.~\ref{fig:profile}).  

Unfortunately, then, the inferred density profile does not judicate between the two central engine models.  A more promising discriminant will come from a spatially resolved location within the host galaxy which can confirm or refute a nuclear spatial location, or the detection of spectro-temporal signatures in the X-ray emission properties which distinguish a stellar-mass from supermassive compact object (e.g., \citealt{Kara+2016}). The host galaxy properties alone may also help discriminate between the two scenarios. In particular, while the hosts of the spectrally-featureless class of optically-selected TDEs prefer to reside near or above the red edge of the green valley in the color-magnitude diagram \citep{hammerstein+2023}, the hosts of GRBs and superluminous SNe are generally bluer star-forming galaxies with low metallicity \citep{Lunnan+2014,Perley+2016}.  Another potential discriminant is the future observation of an abrupt drop in the X-ray light curve as observed in other jetted TDE candidates \citep{Zauderer+2013,Pasham+2015}, which$-$if indeed the result of an accretion disk state-change occurring near the Eddington accretion rate$-$would not be consistent with a stellar-mass compact object. 

\section{Summary}\label{sec:summary}

We have modeled the optical and radio light curves of the luminous transient AT 2022cmc to constrain the properties of the X-ray emitting jet and the density of the surrounding gaseous environment on sub-parsec scales.  Our conclusions can be summarized as follows:
\begin{itemize}
\item Both the radio/mm emission and early optical emission ($t \lesssim 10$ days) are naturally explained by synchrotron emission from a FS blastwave which receives most of its energy during the first few days after the discovery of the transient, consistent with the duration of peak jet activity as implied by the rapidly-evolving internal X-ray emission. 

\item An equipartition analysis reveals that the radio/mm emitting region expands with a Lorentz factors of $\Gamma\gtrsim few$ and energy (in non-thermal electrons and the magnetic field), $E_{\rm eq}\sim10^{49}\,\rm erg$, into an external environment of density profile $k = 1.5-2$, broadly similar to those previously inferred for Swift J1644+57 (Figs.~\ref{fig:equipartition} and \ref{fig:profile}). The roughly constant energy and gradually increasing emitting particle numbers support the FS as the emitting region.

\item The optical peak at 1 day is interpreted as the onset of the self-similar deceleration phase and produced by fast-cooling electrons at the FS. For a narrow jet, the total jet energy is estimated as $E_{\rm j,iso}\sim10^{53}\,\rm erg$ with the density profile and initial Lorentz factor inferred by the equipatition analysis. This coincides with $E_{\rm eq}$ for an assumed beaming factor $\sim10^{-2}$ (jet half-opening angle $\theta_{\rm j} \sim 0.1$) and equipartition parameters of $\varepsilon_{\rm e}\sim0.1$ and $\varepsilon_{\rm B}\sim10^{-3}$.
The observed light curve is fairly well reproduced by the FS afterglow model with these parameters. The implied X-ray radiative efficiency of the jet is high $\varepsilon_{\rm X}\simeq0.6$, also comparable to that of Swift J1644+57 \citep{Berger+2012}. In contrast, a wider spreading jet is disfavored based on energetic arguments. 

\item Unless the jet is highly magnetized, emission from the RS should accompany the FS peak on timescales of $\sim 1$ day.  Though the RS emission was likely not observed in AT 2022cmc, its non-appearance at optical wavelengths places a weak upper limit on the unshocked jet Lorentz factors $\Gamma_{\rm j}\lesssim100$.

\item Despite the similarity of AT 2022cmc in its X-ray/radio/mm emission to other jetted TDE candidates, and its optical plateau phase to some (non-jetted) TDEs, we urge caution in cementing a TDE interpretation for this event.  The location of the transient within its host galaxy has not yet been determined, the short-lived X-ray jet is an order of magnitude smaller than those of other jetted events or the debris fallback timescale for typical TDE SMBH masses $\sim10^{6}\,\Msun$.  The latter tension may be resolved for a more compact star or lower-mass SMBH, but this may be incompatible much longer duration of the highly luminous optical plateau if also powered by fall-back accretion.

\item An alternative model for AT 2022cmc is a core-collapse event of a massive star giving birth to an energetic central engine (magnetar or accreting stellar-mass black hole).  For an active engine timescale of $\sim 10^{5}$ s (spin-down time of the magnetar or fall-back time of the stellar envelope), we demonstrated that a scenario in which roughly half of the engine power goes into the X-ray emitting jet which escapes from the stellar ejecta, and half which thermalizes and boosts the luminosity of the supernova, can explain the entirety of the observations, including in the optical both the early afterglow peak and the later plateau phase (Fig.~\ref{fig:lc3}). 

\item Future observations of the host galaxy and the spatial location of the transient within the host, as well as signatures of an accretion-disk state change in the X-ray light curve, could help distinguish TDE and core-collapse scenarios for AT 2022cmc.  

\end{itemize}

\section*{acknowledgments}
We are grateful to Wenbin Lu for reviewing an early version of the manuscript and providing helpful comments. This work is supported in part by JSPS Overseas Research Fellowships (T.M.).  B.D.M. acknowledges support from the National Science Foundation (AST-2009255).

\section*{data availability}
The data underlying this article will be shared on reasonable request to the corresponding author.

\bibliographystyle{mnras}
\bibliography{reference_matsumoto,ref_Brian}

\label{lastpage}
\end{document}